\documentstyle[prl,aps,epsf]{revtex}

\begin{document}
\bibliographystyle{unsrt}

\draft
\title{ Has the Substructure of Quarks Been Found\\
          by the Collider Detector at Fermilab?}
\author{Keiichi Akama}
\address{Department of Physics, Saitama Medical College,
                     Kawakado, Moroyama, Saitama 350-04, Japan}
\author{Hidezumi Terazawa}
\address{Institute for Nuclear Study, University of Tokyo,
Midori-cho, Tanashi, Tokyo 188, Japan}
\maketitle
\begin{abstract}
The significant excess recently found by the CDF Collaboration 
in the inclusive jet cross section for jet transverse energies 
$E_{T} \ge 200$ GeV over current QCD predictions can be explained 
either by possible production of excited bosons (excited gluons, 
weak bosons, Higgs scalars, etc.) or by that of excited quarks.  
The masses of the excited boson and the excited quark are estimated 
to be around 1600 GeV and 500 GeV, respectively.
\end{abstract}

The CDF Collaboration at the Fermilab Tevatron collider \cite{1} has 
reported their data on the inclusive jet cross section for jet transverse 
energies, $E_{T}$, from 15 to 440 GeV, in the pseudorapidity region, 
$0.1 \le \mid \eta \mid \le 0.7$, with the significant excess over current 
predictions based on perturbative QCD calculations for $E_{T} \ge 200$ GeV, 
which may indicate the presence of quark substructure at the compositeness 
energy scale, $\Lambda_{C}$, of the order of 1.6 TeV.  It can be taken as 
an exciting and already intriguing historical discovery of the substructure 
of quarks (and leptons), which has been long predicted, or as the first 
evidence for the composite models of quarks (and leptons), which has been 
long proposed since the middle of 1970's \cite{2,3,4}.  
Note that such relatively low energy scale for $\Lambda_{C}$ of the order 
of 1 TeV has recently been anticipated rather theoretically \cite{5} or by 
precise comparison between currently available experimental data and 
calculations in the composite models of quarks (and leptons) \cite{6}.  
However, the CDF experimental observation may certainly be taken as a more 
direct evidence for the substructure of quarks.  
The purpose of this letter is to explain the observed excess either by 
possible production of excited bosons (excited gluons, weak bosons, 
Higgs scalars, \underline{etc}.) or by that of excited quarks and to 
estimate the masses of the excited boson and the excited quark to be 
around 1600 GeV and 500 GeV, respectively.

An important motivation for composite models of quarks and leptons is 
to explain the repetition of generation structure in the quark and 
lepton spectrum.  
The repetition of isodoublets of quarks and leptons 
suggests the possible existence of an isodoublet of subquarks, $w_{i} 
(i = 1,2)$ (called ``wakems'' standing for \underline{w}e\underline{ak} 
\underline{e}lectro\underline{m}agnetic), while the repetition of 
color-quartets of quarks and leptons does that of a color-quartet 
of subquarks, $C_{\alpha}$ ($\alpha = 0,1,2,3$) (called ``chroms'' 
standing for colors) \cite{2}.  
Then, the quarks($q$) and leptons($l$) 
 are taken as composites of subquarks including $w_{i}$ and $C_{\alpha}$.
In this picture, the weak bosons ($W^{\pm}$ and $Z$), 
 the gluons ($G^{a}$, $a = 1,2,\cdots,8$), the Higgs scalars $(\phi^{+}, 
 \phi^{0})$ [and even the photon ($\gamma$)], 
 can also be taken as composites of a subquark and an antisubquark 
 such as $w_{i}$ and $\bar{w}_{j}$ or $C_{\alpha}$ and $\bar{C}_{\beta}$.
In these models,
 we expect that there may appear not only exotic states
 and excited states of the fundamental fermions 
 but also those of the fundamental bosons \cite{7}.  
Their expected properties and various effects have been 
 studied extensively in Ref.~\cite{8}.  
In what follows,
 we shall discuss the results of our investigation on 
 the leading-order effects of such excited quarks and bosons 
 to the inclusive jet production cross section for $p \bar{p}$ 
 scattering of $p \bar{p} \rightarrow$ jet + anything.

Let us first consider excited bosons or bosonic composites
 in more general.  
 Let us denote the vector and color-octet, vector and color-singlet, 
 scalar and color-octet, and scalar and color-singlet bosonic composites 
 by $V_{\mu}^{a}$, $V_{\mu}$, $S^{a}$, and $S$, respectively.  
Then, the dimensionless couplings between these bosonic composites 
 and quarks are given by the following interaction Lagrangian:
\begin{eqnarray}
L_{~{\rm \!\! int}~} 
& = & g_{\scriptscriptstyle V8} \overline{q} 
\frac{1}{2} \lambda_{a} {V}_\mu^{a}\gamma^\mu 
({\eta}_{\scriptscriptstyle L8} {\gamma}_{\scriptscriptstyle L} 
+ {\eta}_{\scriptscriptstyle R8} {\gamma}_{\scriptscriptstyle R}) q
+ {g}_{\scriptscriptstyle S8} \overline{q} \frac{1}{2} \lambda_{a} S^{a}q
\nonumber \\
& + & g_{\scriptscriptstyle V1}
 \overline{q} \frac{1}{2} \lambda_{0} {V}_\mu^{a}\gamma^\mu 
(\eta_{\scriptscriptstyle L1} \gamma_{\scriptscriptstyle L} 
+ \eta_{\scriptscriptstyle R1} \gamma_{\scriptscriptstyle R}) q
+ g_{\scriptscriptstyle S1} \overline{q} \frac{1}{2} \lambda_{0} Sq
\label{eq:1}
\end{eqnarray}
where 
 $\gamma_{\scriptscriptstyle L}=(1-\gamma_5)/2$,
 $\gamma_{\scriptscriptstyle R}=(1+\gamma_5)/2$,
 $g_{\scriptscriptstyle V8}$, $g_{\scriptscriptstyle S8}$, 
 $g_{\scriptscriptstyle V1}$,
 $g_{\scriptscriptstyle S1}$ are coupling constants,
 $\lambda_{a}$ ($a=1,2,\cdots, 8$) are the Gell-Mann matrices for color SU(3),
 $\lambda_{0}$ is the $\sqrt{2/3}$ times $3 \times 3$ unit matrix,
 and $(\eta_{\scriptscriptstyle L8},
 \eta_{\scriptscriptstyle R8})$ or $(\eta_{\scriptscriptstyle L1},
 \eta_{\scriptscriptstyle R1}) =$ $(1,1)$, $(1,-1)$, $(1,0)$, and $(0,1)$ 
 for the vector, axial vector, left-handed, and right-handed couplings,
 respectively.  
$V_{\mu}^{a}$ and $V_{\mu}$ are hermitian fields 
 while $S^{a}$ and $S$ are in general complex.  
These interactions respect the chiral symmetry of quarks.  
Note that the dimensionless coupling between gluons,
 $G^{a}$, and $V^{a}$ must have a form of $G^{a\mu \nu} 
 (D_{\mu} V_{\nu}^{a} - D_{\nu} V_{\mu}^{a})$ and,
 therefore, have no physical effect since it can be absorbed into
 the kinetic term of $(G_{\mu \nu}^{a})^{2}$ after diagonalizing
 of $G^{a}$ and $V^{a}$.  
Also note that there exist no dimensionless couplings of $G^{a}$ and $V$,
 $G^{a}$ and $S^{a}$, or $G^{a}$ and $S$.  
Therefore, these bosonic composites contribute to $p \bar{p}$ scatterings
 only through $q \bar{q} \rightarrow q \bar{q}$ scatterings
 and their crossed channels.

Let ($s,t,u$) and $z$ be the Mandelstam variables for the elementary process
 of $q \bar{q} \rightarrow q \bar{q}$ scattering and $\cos{\theta}$ 
 with the scattering angle $\theta$ in the center-of-mass system.  
Then, the differential cross section for the scattering is given by
\begin{eqnarray}
\frac{d \sigma}{dz}
= \frac{1}{36} \cdot \frac{1}{32 \pi s}
\big[A_{L} (s,t,u) + A_{R} (s,t,u) + 2B (s,t,u) + 2B (t,s,u)\big]
\label{eq:2}
\end{eqnarray}
where
\begin{eqnarray}
A_{x} (s,t,u) 
& = & 4u^{2} 
\bigg\{ 2 \mid V_{8}^{xx}(s) \mid^{2}
+2 \mid V_{8}^{xx}(t) \mid^{2}
+9 \mid V_{1}^{xx}(s) \mid^{2}
+9 \mid V_{1}^{xx}(t) \mid^{2}
\nonumber \\
& & \qquad + 2Re
\bigg[- \frac{2}{3} V_{8}^{xx}(s)^{*} V_{8}^{xx}(t) + 4V_{8}^{xx}(s)^{*}
 V_{1}^{xx}(t)
\nonumber \\
& & \qquad \qquad \qquad \qquad
+ 4V_{1}^{xx}(s)^{*} V_{8}^{xx}(t) + 3V_{1}^{xx}(s)^{*} V_{1}^{xx}(t)
\bigg] 
\bigg\},
\ \ \ (x=L,R)
\label{eq:3} 
\\ 
B\ (s,t,u)\  
& = & t^{2} \bigg\{ 
4\big[2 \mid V_{8}^{LR}(s) \mid^{2} + 9 \mid V_{1}^{LR}(s) \mid^{2}\big]
+ t^{2} \big[2 \mid S_{8}(t) \mid^{2} + 9 \mid S_{1}(t) \mid^{2}\big]
\nonumber \\
& & \qquad - 4 Re
\bigg[-\frac{2}{3} V_{8}^{LR}(s)^{*} S_{8}(t) + 4V_{8}^{LR}(s)^{*} S_{1}(t)
\nonumber \\
& & \qquad \qquad \qquad \qquad
+ 4V_{1}^{LR}(s)^{*} S_{8}(t) + 3 V_{1}^{LR}(s)^{*} S_{1}(t)\bigg] \bigg\}
\label{eq:5}
\end{eqnarray}
with the propagators
\begin{eqnarray}
&&
V_{1}^{xy}(s) 
= \cases{\displaystyle
\frac{\displaystyle e^{2}}{\displaystyle s} 
+ \frac{\displaystyle g_{Zx}g_{Zy}}
{\displaystyle s-M_{Z}^{2} + iM_{Z} \Gamma_{Z}}
+ \frac{\displaystyle g_{V1}^{2} \eta_{x1} \eta_{y1}}
{\displaystyle s - M_{V1}^{2} + iM_{V1} \Gamma_{V1}},
& ($x,y = L,R$) \cr
\frac{\displaystyle g_{W}g_{W}'}
{\displaystyle s-M_{W}^{2} + iM_{W} \Gamma_{W}},
& ($x,y=L$) \cr
}
\label{eq:6} 
\\&&
V_{8}^{xy}(s)
= \frac{g^{2}}{s}
+ \frac{g_{V8}^{2} \eta_{x8} \eta_{y8}}{s-M_{V8}^{2} + iM_{V8} \Gamma_{V8}},
\,\,(x,y = L,R)
\label{eq:7} 
\\&&
S_{1}(s) = \frac{g_{S1}^{2}}{s-M_{S1}^{2} + iM_{S1} \Gamma_{S1}},
\label{eq:8} 
\\&&
S_{8}(s) = \frac{g_{S8}^{2}}{s-M_{S8}^{2} + iM_{S8} \Gamma_{S8}}.
\label{eq:9}
\end{eqnarray}
Here, $e$ is the electromagnetic coupling constant,
 $g$ is the QCD coupling constant,
 $g_{\scriptscriptstyle ZL}$ and $g_{\scriptscriptstyle ZR}$
 are the left- and right-handed coupling constants of $Z$ boson,
 $g_{W}$ is the weak gauge coupling constant times the relevant
 CKM matrix element, $M_{X}$ is the mass of particle $X$,
 and $\Gamma_{X}$ is the decay width.  
If the decay of the excited boson is dominated by the two body decay 
 due to the interactions given in Eq. (\ref{eq:1}),
 its decay width is given by
\begin{eqnarray}
&&
\Gamma_{V8} = \Gamma_{V1}
= \frac{M_{V}}{48 \pi} 
\sum{
g_{V}^{2} \sqrt{1-\frac{4m^{2}}{M_{V}^{2}}}
\bigg[\bigg(1-\frac{m^{2}}{M_{V}^{2}}\bigg) (\eta_{L}^{2} + \eta_{R}^{2})
+ \frac{6m^{2}}{M_{V}^{2}} \eta_{L} \eta_{R}\bigg]},
\label{eq:10} 
\\&&
\Gamma_{S8} = \Gamma_{S1}
= \frac{M_{S}}{48 \pi}
\sum{
g_{S}^{2} \sqrt{1-\frac{4m^{2}}{M_{S}^{2}}}
\bigg(1-\frac{2m^{2}}{M_{S}^{2}}\bigg)},
\label{eq:11}
\end{eqnarray}
where $\sum$ denotes the summation over flavors of final quarks
 and $m$'s are the final quark masses,
 all of which but the top quark mass can be practically neglected.

Let us next consider excited quarks (of spin 1/2 for simplicity),
 which are denoted by $Q$'s.  
Then, the interaction of $Q$ with quarks ($q$)
 and gluons ($G_{\mu}^{a}$) is given by
\begin{eqnarray}
L_{int} = -\frac{g_Q}{2M_{Q}}
\bigg[\overline{Q}\frac{1}{2}\lambda^{a}\sigma^{\mu\nu}G_{\mu \nu}^{a} q_{L}
+ \overline{q_{L}} \frac{1}{2} \lambda^{a} \sigma^{\mu \nu} G_{\mu \nu}^{a} Q
+ (L \leftrightarrow R)\bigg]
\label{eq:12}
\end{eqnarray}
where $g_Q$ is a coupling constant and $M_{Q}$ is the excited quark mass. 
Note that an excited quark coupling with $q_{L}$ and another excited quark 
 coupling with $q_{R}$ must be different from one another
 if the chiral symmetry of quarks is preserved.  
If this is the case, the differential cross section for the scattering
 of $q \bar{q} \rightarrow GG$ is given by
\begin{eqnarray}
\frac{d\sigma}{dz}
& = & \frac{1}{27 \pi s}
\bigg[g^{4} (t^{2} + u^{2}) \bigg(\frac{1}{tu} - \frac{9}{4s^{2}}\bigg)
\nonumber \\
& & \qquad \quad +\frac{g^{2} g_Q^{2}}{4M_{Q}^{2}}
Re\bigg(\frac{t^{2}}{t - M_{Q}^{2} + iM_{Q} \Gamma_{Q}} 
+ (t \leftrightarrow u)\bigg)
\nonumber \\
& & \qquad \quad + \frac{g_Q^{4}ut}{M_{Q}^{4}} 
\bigg(\bigg| \frac{t}{t - M_{Q}^{2} + iM_{Q} \Gamma_{Q}} \bigg|^{2} 
+ (t \leftrightarrow u)\bigg)\bigg].
\label{eq:13}
\end{eqnarray}
If the decay of $Q$ is dominated by the interactions
 given in Eq.~(\ref{eq:12}), the decay width of $Q$ is given by
\begin{eqnarray}
\Gamma_{Q} = \frac{g_Q^{2}}{6 \pi} M_{Q}
 \bigg(1 - \frac{m^{2}}{M_{Q}^{2}}\bigg),
\label{eq:14}
\end{eqnarray}
where $m$ is the final quark mass, any one of which except
 the top quark mass can be practically neglected.  
Note that if an excited quark coupling with $q_{L}$
 and another excited quark coupling with $q_{R}$
 are not discriminated against one another,
 which leads to breaking of the chiral symmetry of quarks,
 the above differential cross section would need an additional term,
\begin{eqnarray}
\frac{1}{27 \pi s} \frac{g_Q^{4}ut}{M_{Q}^{4}}
\bigg[\ \bigg| \frac{t}{t - M_{Q}^{2} + iM_{Q} \Gamma_{Q}} \bigg|^{2} 
\ \ \ + \ \ \ (t \leftrightarrow u)\ \ \ \ \ \ \  \ \ \ 
\ \ \ \ \ \ \ \ \ \ \ \  
\nonumber \\
\qquad \qquad \qquad
- \frac{4}{3} Re
\bigg( \frac{tu}{(t-M_{Q}^{2} + iM_{Q} \Gamma_{Q})^{*}(u-M_{Q}^{2}
 + iM_{Q} \Gamma_{Q})}\bigg)\bigg]
\label{eq:15}
\end{eqnarray}
In this case, the decay width of $Q$ would also be changed
 to be twice as much as given in Eq.~(\ref{eq:14}).  
Note also that the differential cross sections for the crossed channels
 can easily be obtained by exchanging ($s,t,u$) appropriately
 and by rewriting the statistical factors due
 to the different spins and colors of initial (and final) quarks (or gluons).

Now we evaluate the single jet $p_{T}$ inclusive distribution, 
 dijet invariant mass distribution, and dijet angular distribution
 in the $p\bar p$ scattering in the CDF energy region.  
For the elementary processes, we take $2 \rightarrow 2$ processes of quarks,
 antiquarks, and gluons.  
Also, we assume that either one of $u$, $d$, $s$, $c$, $b$ quarks
 or gluons in the final states is to be observed as a jet.  
Although the authors of Ref.~\cite{1} have found the excess at high $p_{T}$
 in comparison of their data with the next-to-leading order calculations,
 we have restricted ourselves to the leading order contribution
 from composite models.  
Since higher order corrections are supposed to contribute almost equally
 both in the standard model calculations and in the composite model ones,
 the ratio of the composite model calculation to the standard model
 one may not be so much affected by higher order corrections 
 and may be enough meaningful even if both of the calculations
 are only in the leading order.
As for the parton distribution functions
 we use those of Gl\"{u}ck-Reya-Vogt in Ref.\cite{9}.

In FIG.\ 1, the predictions of the composite models with excited states 
 for the single jet $E_{T}$ inclusive distribution divided 
 by those of the standard model are compared with the 
 corresponding CDF experimental result reported in Ref.\cite{1}. 
We have taken the same average over the pseudorapidity range
 of $0.1 \le \mid \eta \mid \le 0.7$ as the CDF experiment \cite{1}.  
Based on such comparison, we have performed detailed chi-square anlyses,
 and determined the allowed regions (95\% confidence level)
 of the mass $M_{X}$ and coupling constant $\alpha_{X}
 (\equiv g_{X}^{2}/4 \pi)$ of various types of excited 
 states $X$ (See FIG.\ 2). 
It indicates that the excited bosons
 with $\alpha_{X}>0.1$ and $M_{X}>1000$ GeV are allowed.  
The excess of the $E_T$ distribution is well fitted
  by the tail of the high mass resonance of the excited boson.
On the other hand, there is no allowed region for a single excited quark,
 as far as we assume the two-jet decay mode dominates the decay.  
This is because the width (\ref{eq:14}) is too narrow
 to fit the rather gentle slope of the observed data in FIG.~1.  
The width, however, can be broadened due to
 (i) other decay modes such as multi-jet or semi-jet processes,
 (ii) coexistence of several resonances, or
 (iii) limited resolution for the jet energy and momentum measurement.  
Let $r$ be the ratio of the total decay width to the partial 
 width (\ref{eq:14}) of the decay to the two-jet mode.
In FIG.~2, we also show the allowed region
 for the $M_{X}$ and  $\alpha_{X}$
 of the excited quarks for the cases of $r=2$ and 3.  
It is restricted in the low-mass region 400 GeV $< M_{X} <$ 900 GeV
 and $0.03 < \alpha_{X} < 0.8$.

To get more precise information, it may be extremely useful
 to investigate the dijet invariant mass and angular distributions.
Figure 3 shows the predictions with the typical excited states 
 for the dijet invariant mass ($E_{\rm dijet}$) distribution
 divided by those of the standard model.
It predicts a significant excess in the high dijet mass region.
Figure 4(a) shows the predicted dijet angular distribution
 as a function of $\chi \equiv (1+\cos{\theta})/(1-\cos{\theta})$
 (normalized by the average over the region of $1 \le \chi \le 5$).  
The model with excited states predicts relative excess in low $\chi$
 (i.\ e. large $\theta$) region,
 since the peak at $\theta=0$ 
 due to exchange of light quarks and massless gluons  
 is absent in the additional contributions from the excited state.
To see it more clearly, we show in FIG.\ 4(b)
 the ratio of the number of the expected events for $\chi < 2.5$ 
 to that for $\chi > 2.5$ 
 as a function of the dijet invariant mass $E_{\rm dijet}$.  

To sum up, we have shown in this letter that the significant excess found
 by the CDF Collaboration can be explained either by possible production
 of excited bosons whose masses are around 1600 GeV or by that
 of excited quarks whose masses are around 500 GeV.  
The copious production of such excited particles can be expected
 in the future  $e^+e^-$ NLC experiments and $p \bar{p}$ LHC experiments.  
In conclusion, we must mention that although we have assumed the excited
 quarks of spin 1/2 for simplicity, one can also assume those of spin 3/2,
 which has very recently been emphasized by Bander \cite{10}.

\section*{Acknowledgements}
The authors would like to thank Professor Kunitaka Kondo for giving them
 the valuable information on the CDF experiment and for sending them
 the manuscripts before publication.  
One of the authors (H.T.) also wishes to thank Professor Stanley J.~Brodsky
 and all the other staff members, especially Professors James~D.~Bjorken
 and Michael Peskin, of Theoretical Physics Group at Stanford Linear
 Accelerator Center, Stanford University not only for their useful
 discussions on the substructure of quarks but also for their warm
 hospitalities extended to him during his visit in July, 1996 when
 this work was completed.  
The other author (K.A) also wishes to thank Professors Yuichi Chikashige
 and Tadashi Kon for useful discussions and communications.

\def\fcircle {\displaystyle}
\hspace{0mm}
\epsfxsize=14cm\epsffile{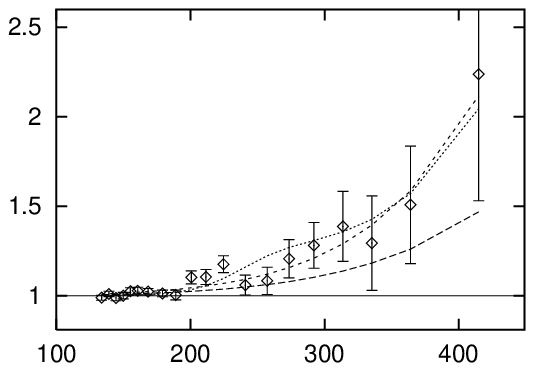}

\begin{picture}(10,10)
\LARGE
\put(110,180){
${\displaystyle{d\sigma \over dE_T}({\rm composite\ model})
\over
  \displaystyle{d\sigma \over dE_T}({\rm standard\ model})}$
}

\put(205,90){$Q$}
\put(250,80){$V8$}
\put(315,80){$V8'$}
\put(340,60){SM}

\put(210,0){$E_T$}
\put(340,0){GeV}
\end{picture}

\def\FC#1{\vskip3mm\noindent{\bf#1}\ \ \ }

\rightskip 20mm
\FC{FIG.~1.}  
The predictions of the composite models with excited states 
 for the single jet $E_{T}$ inclusive distribution divided 
 by those of the standard model.
The label SM indicates the prediction of the standard model,
 $V8$ ($V8'$) indicates that with
 a vector octet excited boson with $M_{V8}$=1600GeV (2000GeV) 
 and $\alpha_{V8}=g_{V8}^2/4\pi$=1, and   
 $Q$ indicates that with the excited quark 
 with $M_{Q}$=500GeV, $\alpha_{Q}=g_Q^2/4\pi$=0.2, and $r$=3,
 where $r$ is the ratio of the decay width to the partial 
 width of the decay to the two-jet mode.
The points with an error bar
 are the  corresponding CDF experimental results reported in Ref.\cite{1}.

\rightskip 0mm

\vspace*{5mm}
\hspace{0mm}
\epsfxsize=14cm\epsffile{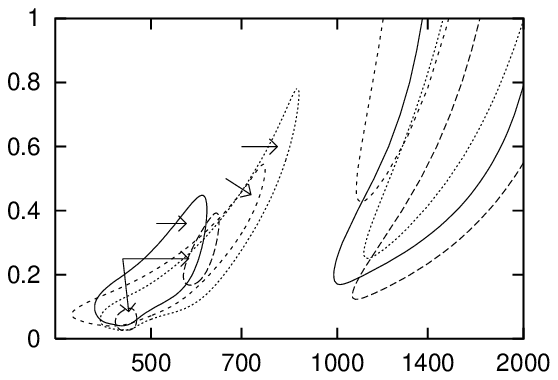}

\begin{picture}(10,10)
\LARGE

\put(235,75){$V8$}
\put(320,75){$V1$}
\put(260,185){$S8$}
\put(345,185){$S1$}
\put(130,170){nonchiral}
\put(160,155){$r=2$}
\put(150,140){$r=3$}
\put( 90,120){chiral}
\put(100,105){$r=3$}
\put( 90,90){$r=2$}
\put(250, 50){excited boson}
\put( 90,200){excited quark $Q$}

\put(30,135){$\alpha_X$}
\put(210,0){$M_X$}
\put(340,0){GeV}
\end{picture}

\rightskip 20mm
\FC{FIG.~2.}  
The allowed regions (95\% confidence level)
 of the mass $M_{X}$ and coupling constant $\alpha_{X}
 (\equiv g_{X}^{2}/4 \pi)$ of various types of excited states $X$. 
 The labels $V8$, $V1$, $S8$, and $S1$ indicate vector-octet,
 vector-singlet, scalar-octet, and scalar-singlet excited bosons,
 respectively.

\rightskip 0mm

\newpage
\hspace{0mm}
\epsfxsize=14cm\epsffile{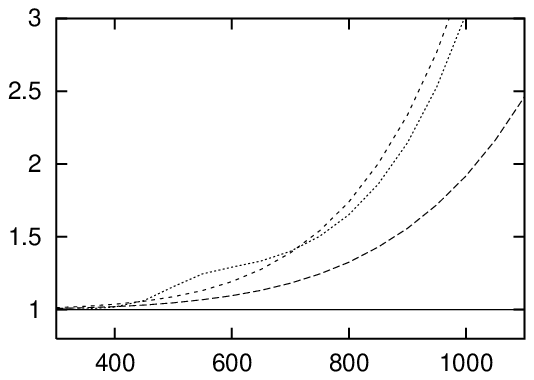}

\begin{picture}(10,10)
\LARGE

\put(90,180){
${\displaystyle{
d\sigma \over dE_{\rm dijet}}({\rm composite\ model})
\over
\displaystyle{
d\sigma \over dE_{\rm dijet}}({\rm standard\ model})}$
}

\put(300,60){SM}
\put(300,210){$V8$}
\put(325,115){$V8'$}
\put(170,90){$Q$}

\put(210,0){$E_{\rm dijet}$}
\put(340,0){GeV}
\end{picture}

\rightskip 20mm
\FC{FIG.~3.}  
The predictions with the typical excited state 
 for the dijet invariant mass ($=E_{\rm dijet}$) distribution
 divided by those of the standard model.
The labels SM, $V8$, $V8'$, and $Q$ are the same as those in FIG.\ 1.

\rightskip 0mm

\vspace*{5mm}
\hspace{0mm}
\hskip60mm
\epsfxsize=8cm\epsffile{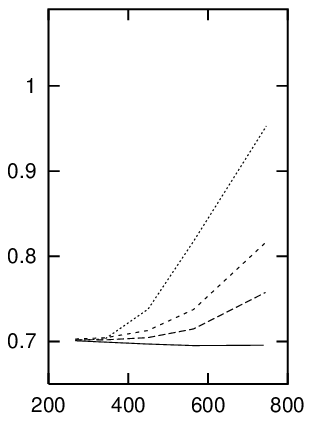}
\hskip-145mm
\epsfxsize=8cm\epsffile{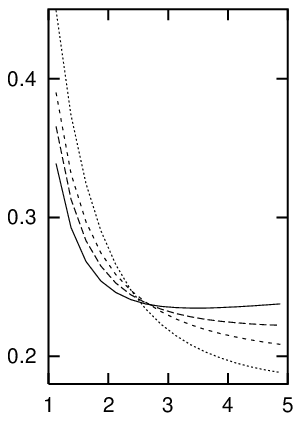}

\begin{picture}(10,10)
\LARGE

\put(25,250){(a)}
\put(210,250){(b)}
\put(90,220){angular}
\put(100,200){distribution}
\put(80,160){$E_{\rm dijet}>625$GeV}
\put(85,125){
$\chi =\fcircle {1+\cos\theta \over 1-\cos\theta }$}
\put(117,0){$\chi $}

\put(130,80){SM}
\put(160,65){$V8'$}
\put(160,50){$V8$}
\put(130,40){$Q$}

\put(250,210){$\displaystyle {
d\sigma (1<\chi <2.5)\over
d\sigma (2.5<\chi <5)}$
}
\put(300,0){$E_{\rm dijet}$}
\put(360,0){GeV}
\put(350,40){SM}
\put(350,65){$V8'$}
\put(350,120){$V8$}
\put(340,165){$Q$}

\end{picture}

\rightskip 20mm
\FC{FIG.~4.}  
(a) The predicted dijet angular distribution
 as a function of $\chi$
 normalized by the average over the region of $1 \le \chi \le 5$.
 (b) The ratio of the number of the expected events for $\chi < 2.5$ 
 to that for $\chi > 2.5$ as a function of 
 the dijet invariant mass $E_{\rm dijet}$.  
The labels SM, $V8$, $V8'$, and $Q$ are the same as those in FIG.\ 1.

\rightskip 0mm

\end{document}